\documentclass[sigconf]{acmart}

\AtBeginDocument{%
  \providecommand\BibTeX{{%
    \normalfont B\kern-0.5em{\scshape i\kern-0.25em b}\kern-0.8em\TeX}}}

\copyrightyear{2025}
\acmYear{2025}
\setcopyright{cc}
\setcctype{by}
\acmConference[CIKM '25]{Proceedings of the 34th ACM International Conference on Information and Knowledge Management}{November 10--14, 2025}{Seoul, Republic of Korea}
\acmBooktitle{Proceedings of the 34th ACM International Conference on Information and Knowledge Management (CIKM '25), November 10--14, 2025, Seoul, Republic of Korea}\acmDOI{10.1145/3746252.3760970}
\acmISBN{979-8-4007-2040-6/2025/11}

\makeatletter
\gdef\@copyrightpermission{
  \begin{minipage}{0.3\columnwidth}
   \href{https://creativecommons.org/licenses/by/4.0/}{\includegraphics[width=0.90\textwidth]{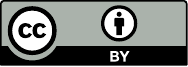}}
  \end{minipage}\hfill
  \begin{minipage}{0.7\columnwidth}
   \href{https://creativecommons.org/licenses/by/4.0/}{This work is licensed under a Creative Commons Attribution International 4.0 License.}
  \end{minipage}
  \vspace{5pt}
}
\makeatother

\usepackage{graphicx}
\usepackage{booktabs}
\usepackage{microtype}
\usepackage{booktabs}
\usepackage{colortbl}
\usepackage{xspace}
\usepackage[most]{tcolorbox}
\usepackage{tcolorbox}
\usepackage{array}
\usepackage{amsfonts}
\usepackage{amsmath}

\usepackage{amssymb}
\usepackage{multirow}
\usepackage{multicol}
\usepackage{listings}
\usepackage{placeins}
\usepackage{subfigure}
\usepackage{graphicx}
\usepackage{enumitem}
\usepackage{hyperref}
\usepackage{tabularx}
\usepackage{pifont}
\newcommand{\imp}[1]{{\small \color[HTML]{34A853} {(#1)}}}

\newcommand{\ours}[1]{BuzzProphet}

\begin{document}

\title{Forecasting the Buzz: Enriching Hashtag Popularity Prediction with LLM Reasoning}

\author{Yifei Xu}
\affiliation{
  \institution{National University of Singapore}
  \country{Singapore}
  }
\email{e1331093@u.nus.edu}
\author{Jiaying Wu}
\authornote{Corresponding authors.}
\affiliation{
  \institution{National University of Singapore}
  \country{Singapore}
  }
\email{jiayingw@nus.edu.sg}
\author{Herun Wan}
\affiliation{
  \institution{Xi'an Jiaotong University}
  \country{Xi'an, China}
  }
\email{wanherun@stu.xjtu.edu.cn}
\author{Yang Li}
\affiliation{
  \institution{Institute of Medical Information, Chinese Academy of Medical Sciences}
  \country{Beijing, China}
  }
\email{li.yang@imicams.ac.cn}
\author{Zhen Hou}
\affiliation{
  \institution{Institute of Medical Information, Chinese Academy of Medical Sciences}
  \country{Beijing, China}
  }
\email{	hou.zhen@imicams.ac.cn}
\author{Min-Yen Kan}
\authornotemark[1]
\affiliation{
  \institution{National University of Singapore}
  \country{Singapore}
  }
\email{kanmy@comp.nus.edu.sg}

\renewcommand{\shortauthors}{Yifei Xu et al.}

\begin{abstract}
Hashtag trends ignite campaigns, shift public opinion, and steer millions of dollars in advertising spend, yet forecasting which tag goes viral is elusive. Classical regressors digest surface features but ignore context, while large language models (LLMs) excel at contextual reasoning but misestimate numbers. We present \textbf{BuzzProphet}, a reasoning-augmented hashtag popularity prediction framework that (1) instructs an LLM to articulate a hashtag’s topical virality, audience reach, and timing advantage; (2) utilizes these popularity-oriented rationales to enrich the input features; and (3) regresses on these inputs. To facilitate evaluation, we release \textbf{HashView}, a 7,532-hashtag benchmark curated from social media. Across diverse regressor—LLM combinations, BuzzProphet reduces RMSE by up to 2.8\% and boosts correlation by 30\% over baselines, while producing human-readable rationales. Results demonstrate that using LLMs as context reasoners rather than numeric predictors injects domain insight into tabular models, yielding an interpretable and deployable solution for social media trend forecasting.\footnote{Data and code are available at: \url{https://github.com/cx-Yifei/BuzzProphet}.}
\end{abstract}

\maketitle

\section{Introduction}

\begin{figure}[!t]
  \centering
  \includegraphics[width=0.8\linewidth]{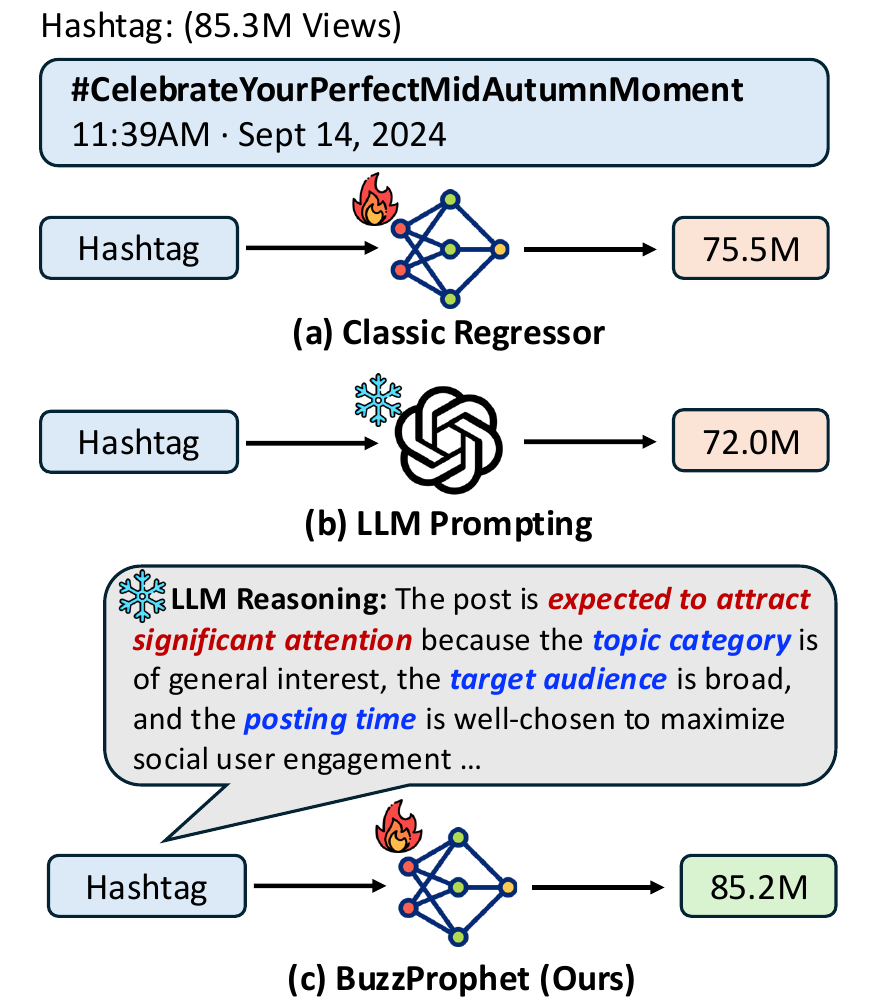}
\caption{Comparison of \ours{} with prior work.}
  \label{fig:intro}
\end{figure}

Hashtags, short textual strings preceded by the hash symbol (\#), are among the most effective tools for enhancing visibility and engagement on social media \cite{bruns2015twitter,la2022research}. Accurately predicting their future reach (e.g., view counts) is valuable for both content creators seeking to optimize engagement \cite{chakrabarti2023hashtag} and platform managers allocating resources and anticipating cascades \cite{laestadius2017mobilizing}. Unlike posts, hashtags serve as cross-cutting topical markers that transcend accounts and timelines, facilitating discovery, trend formation, and collective discourse—making their popularity more volatile yet also more informative for forecasting.

Despite this importance, hashtag popularity has received limited attention. Prior work treated it as a \textit{classification} task \cite{ma2012will}, using arbitrary thresholds to bucket popularity. A more principled framing is \textit{regression}, where the goal is to estimate a scalar popularity score, often log-normalized view count \cite{wu2023smp}. Existing approaches \cite{ding2019social,wu2022deeply} typically employ regressors such as CatBoost \cite{prokhorenkova2018catboost}, but these fail to capture rich contextual signals. Large language models (LLMs) offer strong interpretive and reasoning capabilities for social media content \cite{feng2024bot,zeng2024large}, yet remain unreliable for direct numeric prediction \cite{chen2024clinicalbench,yang2025number}, often yielding inconsistent estimates \cite{huang2025survey,xiong2024can}.

Motivated by recent advances in enriching smaller models with LLM-generated rationales \cite{he2024harnessing,hu2024bad}, we propose \textbf{\ours{}}, a \textit{reasoning-augmented} regression framework combining the numeric precision of classical regressors with the contextual reasoning of LLMs. As shown in Figure~\ref{fig:intro}, \ours{} leverages LLMs to generate textual rationales on three key engagement factors: \textbf{(1) topic category} \cite{grover2022evolution}, \textbf{(2) target audience} \cite{fan2023influencers,van2024less}, and \textbf{(3) posting time} \cite{shahbaznezhad2021role,trunfio2021conceptualising}. These free-form explanations are encoded and fused into a regression model, yielding more accurate and stable predictions.

To support systematic evaluation, we further introduce \textbf{HashView}, a large-scale benchmark consisting of 7,532 hashtags from diverse domains, curated from influential accounts on Weibo\footnote{\url{https://weibo.com/}} (the Chinese equivalent of X). Extensive experiments demonstrate that \ours{} consistently outperforms strong regression and LLM baselines, while also providing interpretable reasoning that reveals why certain hashtags are likely to attract attention. Our findings highlight the value of combining symbolic reasoning with statistical estimation for context-aware, explainable trend forecasting.

\section{Related Work}
Predicting social media engagement has practical value in content optimization \cite{agarwal2008online}, recommendation \cite{chakrabarti2023hashtag}, and moderation \cite{laestadius2017mobilizing}. Existing work falls into two categories: (1) predicting content popularity (e.g., views, likes) from post snapshots at posting time \cite{ding2019social,hu2024revisit,tu2024higher,wu2022deeply,wu2023smp}, and (2) modeling topic trends from historical engagement data \cite{xu2025smtpd,yang2011patterns}. Most approaches extract multimodal features with pre-trained models (e.g., ResNeXt-101 \cite{xie2017aggregated}, BERT \cite{devlin2019bert}) and apply regressors like CatBoost \cite{prokhorenkova2018catboost}. Hashtags, despite their  role in signaling and aggregating attention, are understudied. Early work \cite{ma2012will} treated hashtag popularity prediction as classification with arbitrary thresholds. To enable more precise modeling and capture subtle differences in engagement levels, we instead frame it as fine-grained regression, introducing a reasoning-augmented framework with a benchmark tailored to hashtags.

LLMs exhibit emergent abilities in analyzing social media content \cite{kmainasi2025llamalens,peters2024large,yang2024mentallama} and simulating user behaviors online \cite{gao2023s3,nan2024let,tornberg2023simulating}. Yet, prior studies highlight their limitations in numeric prediction \cite{chen2024clinicalbench,yang2025number} and their tendency to yield inconsistent outputs \cite{huang2025survey,xiong2024can}, restricting direct use in popularity forecasting. To our knowledge, we are the first to adapt LLMs for this task by coupling the numeric precision of regressors with the contextual reasoning of LLMs.

\section{Problem Formulation}
\label{sec:formulation}

Given a social media hashtag, the task is to predict its future popularity, measured by view counts. Each instance is represented as a pair $(x, t)$, where $x$ is the textual content and $t$ is the posting time. The prediction target is a scalar $y$ denoting popularity. To address the high variance of view counts and stabilize regression, we follow prior work \cite{wu2023smp} and apply log normalization, training models to predict $\log(y+1)$.

\section{Our Approach: \ours{}}
\label{sec:approach}

We present \ours{}, a reasoning-augmented regression framework for hashtag popularity prediction. Inspired by recent advances in leveraging LLM-generated explanations to enhance smaller models across domains \cite{he2024harnessing,hu2024bad}, \ours{} bridges the gap between the numeric precision of classical regressors and the social reasoning of large language models, offering a practical yet interpretable solution (as overviewed in Figure~\ref{fig:intro}). 

\underline{\textbf{Popularity-Oriented Reasoning Elicitation.}}
At the core of \ours{} is the elicitation of human-readable rationales that reflect why a hashtag may or may not become popular. Rather than relying solely on surface-level hashtag textual features, we prompt an LLM $\mathcal{M}{r}$ to generate explanatory rationales $r = \mathcal{M}{r}(x, t)$ based on the hashtag content $x$ and its posting time $t$. We focus on three dimensions that are empirically and theoretically grounded in influencing social media visibility:
\textbf{(1) Topic Category}: certain topics (e.g., breaking news, celebrity gossip) are inherently more viral \cite{grover2022evolution};
\textbf{(2) Target Audience}: content tailored to specific user segments or broadly resonant across diverse groups tends to elicit higher engagement \cite{fan2023influencers,van2024less};
\textbf{(3) Posting Time}: temporal context affects visibility due to user activity patterns and algorithmic promotion \cite{shahbaznezhad2021role,trunfio2021conceptualising}. As illustrated in Figure~\ref{fig:case_study}, for each dimension, the LLM first (1) makes a \textit{prediction} (e.g., ``the topic is about entertainment'') and then (2) provides an \textit{explanation} of its expected impact on popularity (e.g., ``entertainment-related posts are likely to attract significant attention''). Finally, the LLM produces an overall summary that synthesizes these factors (e.g., ``moderate-high final popularity''). This structured reasoning transforms implicit contextual cues into explicit, interpretable signals, enabling downstream regressors to exploit high-level insights that are otherwise inaccessible from the raw hashtag text alone.

\underline{\textbf{Reasoning-Enriched Hashtag Encoding.}} To incorporate these popularity-related insights, we construct an enriched textual input by concatenating the original hashtag text $x$, its posting time $t$, and the generated reasoning $r$:
\begin{equation}
x_{\text{aug}} = \textsf{CONCAT}(x, t, r).
\end{equation}
We then encode $x_{\text{aug}}$ with a pre-trained LM $\mathcal{M}_{\text{emb}}$ (e.g., RoBERTa~\cite{liu2019roberta}) with frozen parameters to obtain a dense semantic representation $\mathbf{h} = \mathcal{M}_{\text{emb}}(x_{\text{aug}}) \in \mathbb{R}^{D}$.
This enables the model to capture both content-level semantics and auxiliary LLM-inferred popularity signals in a unified embedding space, facilitating downstream training.

\underline{\textbf{Popularity Prediction.}}
Finally, we feed the reasoning-enriched representation $\mathbf{h}$ into a lightweight regression model $g_\phi$ (e.g., CatBoost~\cite{prokhorenkova2018catboost}) to predict the expected hashtag popularity score:
\begin{equation}
\hat{y} = g_\phi(\mathbf{h}).
\end{equation}

By using classical regressors as model backbone, \ours{} benefits from stable numeric prediction, while the reasoning component injects high-level contextual understanding. This modular design also ensures interpretability and ease of deployment, as the reasoning component can be updated or ablated independently.

\section{Experiments}

We evaluate both the quantitative prediction performance and qualitative interpretability of \ours{}. To enable systematic study, we introduce \textbf{HashView}, a large-scale benchmark for hashtag popularity prediction. Curated from Weibo, HashView contains 7,532 hashtags that trended between September and November 2024. For each hashtag, we record its text, posting time, and cumulative view count (with 99.5\% of final counts occurring within 23.2 hours after posting), yielding a reliable measure of early popularity. The prediction target is the log-normalized view count (see Section~\ref{sec:formulation}).

\begin{figure}[!t]
  \centering
  \includegraphics[width=0.95\linewidth]{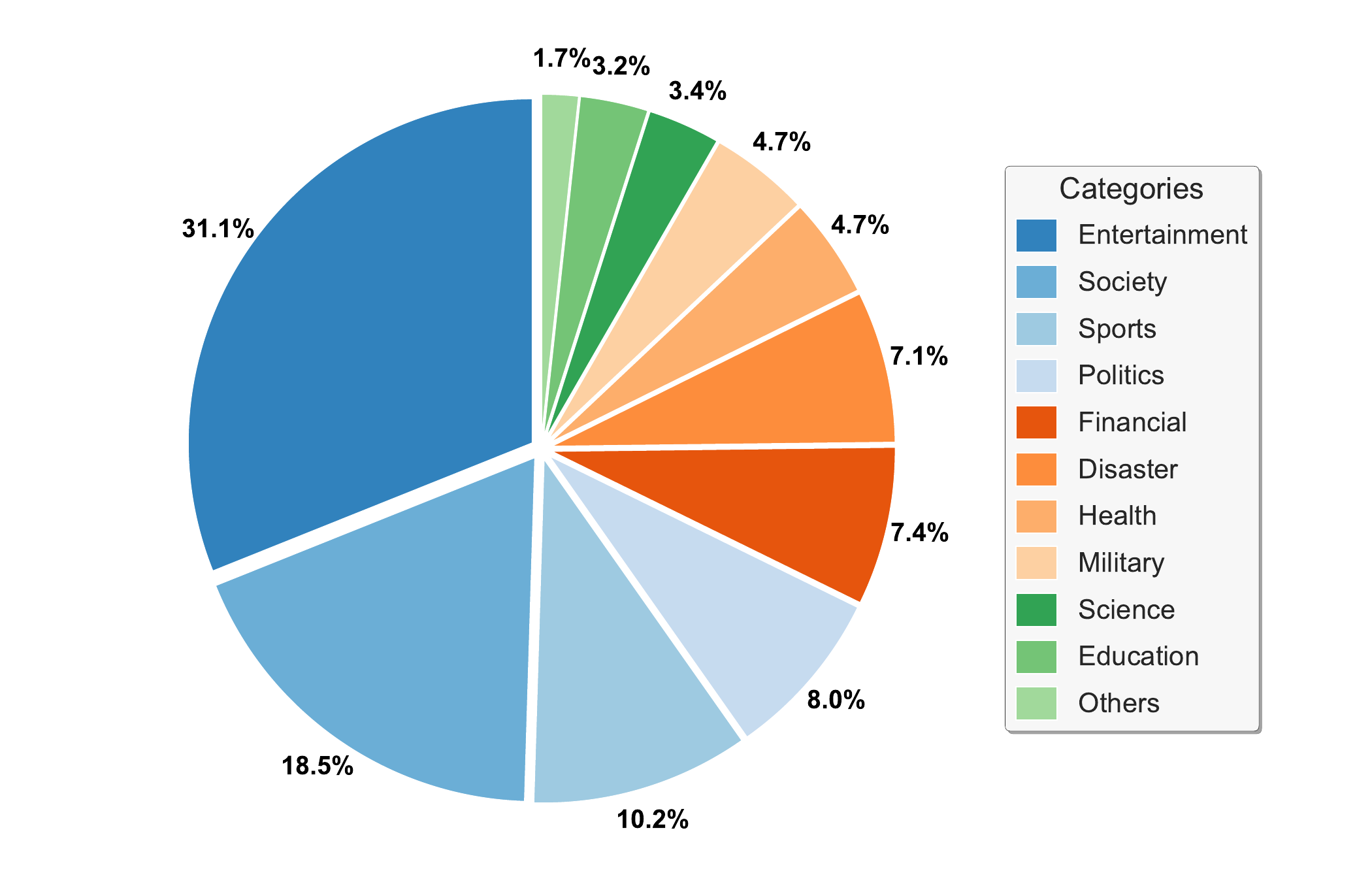}
  \caption{Domain distribution of our HashView benchmark.}
  \label{fig:topic-dist}
\end{figure}

\begin{figure}[!t]
  \centering
  \includegraphics[width=1.0\linewidth]{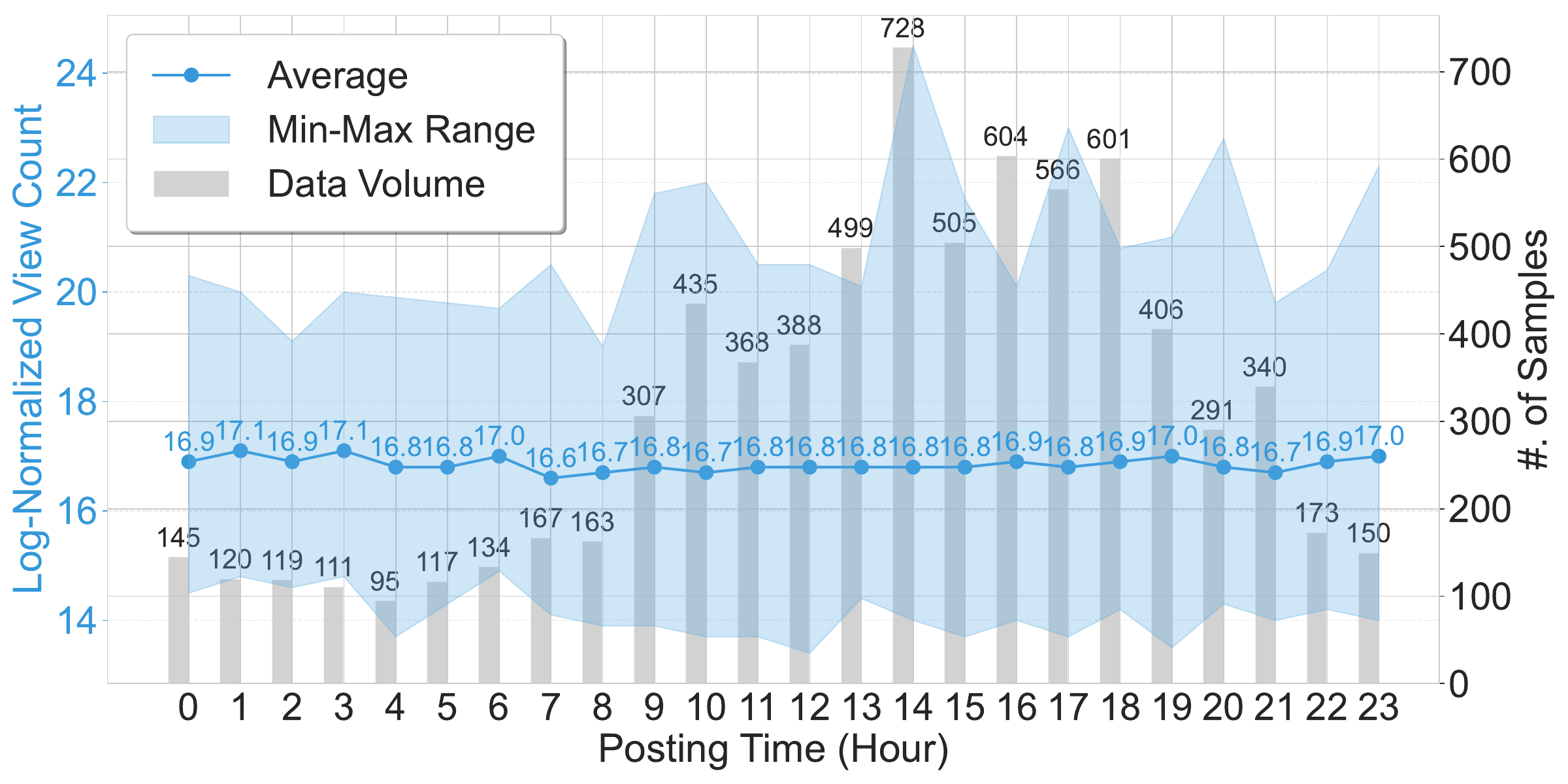}
\caption{Temporal distribution of hashtag postings in HashView, bucketized by hour of day.} 
\label{fig:temporal-trend}
\end{figure}

Figure~\ref{fig:topic-dist} shows the topical distribution of hashtags, while Figure~\ref{fig:temporal-trend} depicts temporal posting and view count patterns. To better reflect real-world deployment, we adopt a \textit{temporal split}, dividing the dataset chronologically into the first 80\% for training and the last 20\% for testing.  This mirrors practical scenarios where models learn from historical data to forecast future trends.

\subsection{Experimental Setup}
\label{sec:exp-setup}
\underline{\textbf{Baselines.}} We compare against two categories: \textbf{(1) LLM few-shot prediction.} Each model is prompted to output a scalar value for the (log-transformed) view count, constrained to the observed range in HashView. Prompts include one demonstration per view-level bucket (e.g., 14.5, 15.8, 16.4). We evaluate three LLMs: Llama3-8B-Chinese-Chat \cite{wang2024llamacn} (open-source LLM), GPT-4o \cite{hurst2024gpt} (proprietary LLM), and GPT-o3-mini \cite{openai2025o3mini} (proprietary Large Reasoning Model). \textbf{(2) Classical regression models.} We implement RandomForest \cite{liaw2002rf}, LightGBM \cite{ke2017lightgbm}, and CatBoost \cite{prokhorenkova2018catboost}, covering both bagging and boosting paradigms widely applied in structured prediction.

\underline{\textbf{Implementation Details.}} Llama3-8B was run on two A40 GPUs; GPT-4o (\texttt{gpt-4o-2024-08-06}) and o3-mini (\texttt{o3-mini-2025-01-31}) were accessed via OpenAI API. We used PyTorch 2.4.0, Transformers 4.46.3, and Scikit-learn 1.3.2, with a fixed random seed of 42. We extract 768-dimensional hashtag features via a frozen multilingual XLM-RoBERTa model (\texttt{xlm-roberta-base}) \cite{conneau2020unsupervised}. Hyperparameters of regressors were tuned using \texttt{RandomizedSearchCV} ($n_{\text{iter}}{=}30$, 3-fold CV). Key search spaces for classical regressors and \ours{} include the following. (1) RandomForest: estimators $\in$ [100, 500]; (2) LightGBM/CatBoost: learning rate $\in$ log-uniform[0.001, 0.1]; (3)MLP: learning rate $\in$ log-uniform[0.001, 0.01], 1-3 layers, batch size $\in$ \{32, 64, 128, 256\}.

\underline{\textbf{Evaluation Metrics.}} 
We adopt three widely used regression metrics: Root Mean Squared Error (RMSE), Mean Absolute Error (MAE), and Spearman's Rank Correlation (SRC). 

\subsection{Results}
\label{sec:results}

\underline{\textbf{Popularity Prediction Performance.}}
We compare \ours{} with classical regressors and LLM-only few-shot baselines across multiple configurations. Table~\ref{tab:main-results} reveals two key findings: \textit{(1) LLMs perform poorly when used directly for numeric prediction}, indicating that strong language understanding alone does not yield accurate popularity estimates. \textit{(2) Incorporating LLM reasoning into regressors consistently improves performance}, with gains tied to reasoning quality: weaker models like Llama3-8B provide little benefit, while stronger GPT-4o and o3-mini models substantially improve accuracy. These results validate the importance of high-quality reasoning augmentation in popularity prediction.

\begin{table}[t]
\centering
\caption{Hashtag popularity prediction performance. Best and second-best results are bolded or underlined. Parenthesized percentages give relative improvements over the corresponding regressor baselines. For brevity, Llama3-8B refers to \texttt{Llama3-8B-Chinese-Chat} \cite{wang2024llamacn}, a model instruction-tuned on Chinese corpora.}
\label{tab:main-results}
\resizebox{\columnwidth}{!}{
\begin{tabular}{lccc} 
    \toprule
    \textbf{Method} & \textbf{RMSE $\downarrow$} & \textbf{MAE $\downarrow$} & \textbf{SRC $\uparrow$} \\
    \midrule
    \multicolumn{4}{c}{\cellcolor[HTML]{EFEFEF}\textbf{\textit{Few-Shot LLM Prompting}}} \\
    Llama3-8B & 1.697 & 1.222 & 0.017 \\
    GPT-4o & 1.599 & 1.290 & 0.079 \\
    GPT-o3-mini & 1.813 & 1.494 & 0.065 \\
    \midrule 
    \textbf{RandomForest} & 1.053 & 0.836 & 0.300 \\
    \quad + \ours{} (Llama3-8B) & 1.063 & 0.846 & 0.286 \\
    \quad + \ours{} (GPT-4o) & \underline{1.031 \imp{2.09\%}} & \underline{0.811 \imp{2.99\%}} & \underline{0.360 \imp{20.00\%}} \\
    \quad + \ours{} (o3-mini) & \textbf{1.024 \imp{2.75\%}} & \textbf{0.803 \imp{3.95\%}} & \textbf{0.387 \imp{29.00\%}} \\
    \midrule
    \textbf{LightGBM} & 1.058 & 0.835 & 0.277 \\
    \quad + \ours{} (Llama3-8B) & 1.057 & 0.841 & 0.284 \imp{2.53\%} \\
    \quad + \ours{} (GPT-4o) & \underline{1.039 \imp{1.80\%}} & \underline{0.818 \imp{2.04\%}} & \underline{0.339 \imp{22.38\%}} \\
    \quad + \ours{} (o3-mini) & \textbf{1.028 \imp{2.83\%}} & \textbf{0.809 \imp{3.11\%}} & \textbf{0.361 \imp{30.32\%}} \\
    \midrule
    \textbf{CatBoost} & 1.035 & 0.821 & 0.332 \\
    \quad + \ours{} (Llama3-8B) & 1.032 \imp{0.29\%} & 0.827 & 0.354 \imp{6.63\%} \\
    \quad + \ours{} (GPT-4o) & \underline{1.020 \imp{1.45\%}} & \underline{0.803 \imp{2.19\%}} & \underline{0.380 \imp{14.46\%}} \\
    \quad + \ours{} (o3-mini) & \textbf{1.018 \imp{1.64\%}} & \textbf{0.802 \imp{2.31\%}} & \textbf{0.387 \imp{16.57\%}} \\
    \midrule
    \textbf{MLP Regressor} & 1.090 & 0.863 & 0.274 \\
    \quad + \ours{} (Llama3-8B) & 1.089 \imp{0.09\%} & 0.868 & 0.280 \imp{2.19\%} \\
    \quad + \ours{} (GPT-4o) & \underline{1.083 \imp{0.64\%}} & \underline{0.854 \imp{1.04\%}} & \underline{0.298 \imp{8.76\%}} \\
    \quad + \ours{} (o3-mini) & \textbf{1.061 \imp{2.66\%}} & \textbf{0.836 \imp{3.13\%}} & \textbf{0.329 \imp{20.07\%}} \\
    \bottomrule
\end{tabular}}
\end{table}

\underline{\textbf{Ablation Study.}}
We evaluate the contribution of each reasoning dimension via ablations, using \ours{} variants that isolate a single popularity-oriented rationale generated by o3-mini. Table~\ref{tab:ablation} shows three key observations: \textit{(1) Topic category and target audience reasoning each substantially improve prediction.} \textit{(2) Posting-time reasoning improves SRC but offers limited benefit in RMSE and MAE}, likely because time is already an explicit input feature; as Figure~\ref{fig:temporal-trend} shows, temporal patterns are highly correlated with popularity, leaving limited signal for LLM rationales. \textit{(3) The full \ours{} model, combining all three reasoning dimensions, achieves the best results}, confirming the value of multi-faceted reasoning for modeling hashtag popularity.

\begin{table}[t]
\centering
\caption{Ablation study of \ours{} through variants that reason about a single popularity indicator.}
    \label{tab:ablation}
\resizebox{\columnwidth}{!}{
    \begin{tabular}{lccc} 
        \toprule
        \textbf{Method} (w/ o3-mini Reasoning) & \textbf{RMSE $\downarrow$} & \textbf{MAE $\downarrow$} & \textbf{SRC $\uparrow$} \\
        \midrule
        \textbf{RandomForest} & 1.053 & 0.836 & 0.300 \\
        \quad + Topic Category & \underline{1.031} & \underline{0.808} & \underline{0.374} \\
        \quad + Target Audience & 1.032 & 0.810 & 0.356 \\
        \quad + Posting Time & 1.055 & 0.832 & 0.320 \\
        \quad + \ours{} (All Dimensions) &  \textbf{1.024} & \textbf{0.803} & \textbf{0.387} \\
        \midrule
        \textbf{MLP Regressor} & 1.090 & 0.863 & 0.274 \\
        \quad + Topic Category  & \underline{1.068} & \underline{0.849} & \underline{0.318} \\
        \quad + Target Audience & 1.079 & 0.860 & 0.289 \\
        \quad + Posting Time & 1.093 & 0.866 & 0.280 \\
        \quad + \ours{} (All Dimensions) & \textbf{1.061} & \textbf{0.836} & \textbf{0.329} \\
        \bottomrule
    \end{tabular}
}
\end{table}

\underline{\textbf{Effects of Hashtag Encoders: LM vs. LLM.}} 
To test whether LLM embeddings provide richer semantic representations for hashtag content \cite{tang2024understanding}, we compare regressors using features from a pre-trained LM (xlm-roberta-base) and an LLM (Llama3-8B). Table~\ref{tab:encoder} shows that Llama-based features consistently underperform RoBERTa-based ones. A likely reason is that hashtags are short and syntactically simple, offering little advantage from high-dimensional LLM representations. These findings suggest that lightweight LMs remain strong, efficient, and well-suited encoders for hashtag popularity prediction.
\begin{table}[t]
\centering
\caption{Prediction performance of regressors using LM (RoBERTa-base) vs. LLM (Llama3-8B) as hashtag encoders.}
\label{tab:encoder}
\begin{tabular}{llccc}
\toprule
\textbf{Method} & \textbf{Encoder} & \textbf{RMSE $\downarrow$} & \textbf{MAE $\downarrow$} & \textbf{SRC $\uparrow$} \\
\midrule
\multirow{2}{*}{RandomForest} & RoBERTa & \textbf{1.053} & \textbf{0.836} & \textbf{0.300} \\
& Llama3-8B & 1.068 & 0.847 & 0.247 \\
\midrule
\multirow{2}{*}{LightGBM}     & RoBERTa & \textbf{1.058} & \textbf{0.835} & \textbf{0.277} \\
& Llama3-8B & 1.061 & 0.839 & 0.264 \\
\midrule
\multirow{2}{*}{CatBoost}     & RoBERTa & \textbf{1.035} & 0.821 & \textbf{0.332} \\
& Llama3-8B & 1.040 & \textbf{0.819} & 0.322 \\
\bottomrule
\end{tabular}
\end{table}

\begin{figure}[!t]
  \centering
  \includegraphics[width=0.9\linewidth]{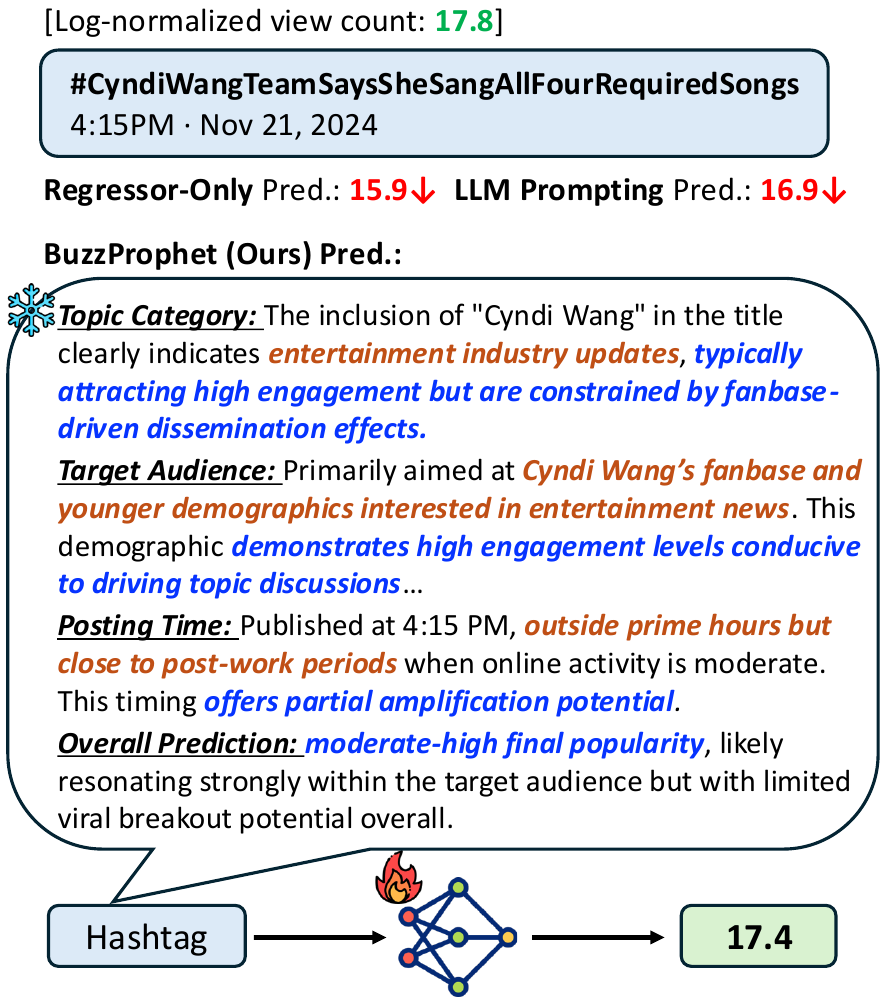}
\caption{Illustration of how \ours{} generates more accurate predictions through interpretable reasoning. (Orange: LLM predictions for the three dimensions; blue: explanations about their potential influence on hashtag popularity.)}
  \label{fig:case_study}
\end{figure}

\underline{\textbf{Qualitative Case Study.}}
In Figure~\ref{fig:case_study}, we show how \ours{} presents human-readable reasoning traces that enhance both prediction effectiveness and interpretability. By decomposing hashtag popularity into predictions and explanations across three popularity-indicative dimensions, \ours{} provides valuable insights that guide the regression model toward more accurate estimations.

\section{Conclusion}
We introduce \ours{}, the first reasoning-augmented framework that couples the contextual reasoning of LLMs with the numeric precision of classical regressors for social media hashtag popularity prediction. On our newly proposed HashView benchmark, \ours{} consistently enhances both predictive accuracy and interpretability across diverse regressor–LLM pairings, demonstrating strong adaptability. These results highlight the value of positioning LLMs not as direct predictors, but as reasoning engines that enrich downstream social media analytics with structured, human-readable insights.

\underline{\textbf{Limitations and Future Work.}} 
While \ours{} establishes a novel and effective paradigm, several avenues remain open for future exploration. HashView is currently limited to Chinese data from Weibo; extending it to other platforms (e.g., X/Twitter) and languages would better assess generalizability. Moreover, our evaluation excludes graph-based baselines such as repost-cascade modeling, due to the lack of hashtag-level dissemination data; incorporating such relational signals when feasible could offer further gains. Pursuing these directions would broaden both the applicability and the impact of reasoning-augmented popularity prediction.

\begin{acks}
This research was supported by NUS School of Computing Large Project Development Seed Grant (A-8002749-00-00) and Chinese Academy of Medical Sciences Innovation Fund for Medical Sciences (2021-I2M-1-033). 
\end{acks}

\section*{GenAI Usage Disclosure}
Generative AI was used only to generate auxiliary reasoning features within the proposed framework and for minor grammatical editing, with human oversight ensuring integrity, accuracy, and accountability throughout the research process.

\bibliographystyle{ACM-Reference-Format}
\bibliography{references}

\end{document}